\documentclass[conference]{IEEEtran}
\IEEEoverridecommandlockouts
\usepackage{cite}
\usepackage{algorithmic}
\usepackage{array}
\usepackage{textcomp}
\usepackage{tikz}
\usepackage{xcolor}
\usepackage{pgfplots}
\usepackage{pgfplotstable}
\usepackage{amsmath,amssymb,amsfonts}
\usepackage{cases}
\usetikzlibrary{shadows}
\usetikzlibrary{patterns}
\usetikzlibrary{arrows,arrows.meta,positioning,shadows,shapes,backgrounds,calc,fit}

\def\BibTeX{{\rm B\kern-.05em{\sc i\kern-.025em b}\kern-.08em
    T\kern-.1667em\lower.7ex\hbox{E}\kern-.125emX}}

\begin{document}
\title{On Parameter Optimization and Reach Enhancement for the Improved Soft-Aided Staircase Decoder\\
\thanks{This work is partially supported by the NSFC Program (No. 62001151) and Fundamental Research Funds for the Central Universities (JZ2020HGTA0072, JZ2020HGTB0015), and Anhui Provincial Natural Science Foundation (2008085QF282).
The work of G. Liga is funded by the EUROTECH postdoc programme under the European Union’s Horizon 2020 research and innovation programme (Marie Skłodowska-Curie grant agreement No 754462). The work of A. Alvarado is supported by the Netherlands Organisation for Scientific Research (NWO) via the VIDI Grant ICONIC (project number 15685) and the European Research Council (ERC) under the European Unions Horizon 2020 research and innovation programme (grant agreement No. 757791).}
}

\author{\IEEEauthorblockN{Yi Lei, Bin Chen}
\IEEEauthorblockA{\textit{Dept. of Computer Science and Information Engineering }\\
\textit{ (Hefei University of Technology)}\\
Hefei, China\\
\{leiyi,bin.chen\}@hfut.edu.cn}\vspace{-1em}
\and
\IEEEauthorblockN{Gabriele Liga, Alex Alvarado}
\IEEEauthorblockA{\textit{Dept. of Electrical Engineering}\\
\textit{ (Eindhoven University of Technology)}\\
Eindhoven, The Netherlands \\
\{g.liga,a.alvarado\}@tue.nl}
\vspace{-1em}}

\maketitle
\begin{abstract}
The so-called improved soft-aided bit-marking algorithm was recently proposed for staircase codes (SCCs) in the context of fiber optical communications. This algorithm is known as iSABM-SCC. With the help of channel soft information, the iSABM-SCC decoder marks bits via thresholds to deal with both miscorrections and failures of hard-decision (HD) decoding.
In this paper, we study iSABM-SCC focusing on the parameter optimization of the algorithm and its performance analysis, in terms of the gap to the achievable information rates (AIRs) of HD codes and the fiber reach enhancement. We show in this paper that the marking thresholds and the number of modified component decodings heavily affect the performance of iSABM-SCC, and thus, they need to be carefully optimized. By replacing standard decoding with the optimized iSABM-SCC decoding, the gap to the AIRs of HD codes can be reduced to $0.26$--$1.02$~dB for code rates of $0.74$--$0.87$ in the additive white Gaussian noise channel with $8$-ary pulse amplitude modulation. The obtained reach increase is up to $22\%$ for data rates between $401$~Gbps and $468$~Gbps in an optical fiber channel.
\end{abstract}

\vspace{-0.5em}
\section{Introduction}

As targeted data rates exceed $400$~Gbps, simple but powerful hard-decision (HD) forward-error-correction (FEC) codes has gained renewed interest in optical transport networks (OTNs). Staircase codes (SCCs)~\cite{Smith2012}, which use iterative bounded-distance-decoding (BDD), is a family of HD-FEC codes that is particularly popular in optical communications. So far, SCCs have been recommended in several OTN standards, e.g., ITU-T G.709.2/Y.133.2~\cite{G709.2}, G.709.3/Y.133.3~\cite{G709.3} and OIF 400ZR Implementation Agreement~\cite{OIF400G}.
Although SCCs can achieve quite high coding gains (compared to other HD-FEC codes), there still exists a significant performance loss with respect to the achievable information rate (AIR) of HD-FEC as well as to the AIR of soft-decision (SD) FEC. For this reason, the design of improved SCC decoders is a hot topic in recent years, with a particular focus on good performance-complexity tradeoffs.

As each bit in SCCs is protected by two component codewords, one idea for improving the performance of SCCs is to prevent miscorrections by checking conflicts between the two component codewords~\cite{Christian1} or solving so-called \emph{stall patterns} via bit flipping (BF) through the intersections of nonzero-syndrome component words~\cite{Holzbaur2017}. Although these methods are simple as they only operate on binary messages, their gains are limited. To obtain a higher gain, an extreme solution is to completely replace the BDD component decoder with an SD decoder~\cite{Douxin_ISTC2018,CondoOFC2020}. However, this solution greatly increases decoding complexity.

Recently, a new class of decoding algorithms---often called soft-aided HD (SA-HD) decoders---have been shown to provide a good compromise between complexity and performance. The main idea is to \emph{assist} the HD decoding by using channel soft information, i.e., log-likelihood ratios (LLRs). For example,~\cite{Alireza} proposed to make a hard decision based on the weighted sum of the BDD output and the channel LLR, while~\cite{AlirezaarXiv2019} replaced BDD with generalized minimum distance decoding in which the erasures are determined via channel LLRs. Lately, an extension work of~\cite{Alireza}, called BEE-SCC, is reported in~\cite{AlirezaSCC2020}. BEE-SCC refined the reliability combining rule via density evolution and performed an extra decoding attempt based on error and erasure decoding of the component codes.

In~\cite{YiISTC2018,YiTCOM2019}, a soft-aided bit-marking (SABM) algorithm was proposed for SCCs. We will refer to this algorithm as SABM-SCC. Differently from the methods proposed in~\cite{Alireza,AlirezaarXiv2019,AlirezaSCC2020}, the soft information in SABM-SCC is only used to mark bits as highly reliable bits (HRBs) and highly unreliable bits (HUBs). The marked bits are then used to deal with both miscorrections and failures of BDD. Although the SABM-SCC decoder was shown to achieve considerable gains, its implementation is not straightforward, as sorting bits by reliability is required for every row of an SCC block for marking HUBs. With the idea of targeting hardware-friendly implementations, an improved SABM (iSABM) algorithm was recently introduced for SCCs (called iSABM-SCC)~\cite{Yi_iSABM2021}. The iSABM-SCC decoder uses two thresholds to classify the bits into HRBs, HUBs and uncertain bits (UBs). Combined with random selection of HUBs for bit flipping, gains up to $0.91$~dB with respect to standard SCCs were reported~\cite{Yi_iSABM2021}. The achieved gain is slightly higher than that of the state-of-the-art BEE-SCC decoder proposed in~\cite{AlirezaSCC2020}, however, a thorough parameter optimization and characterization of the iSABM-SCC decoder is yet to be performed.

In this paper, we start by reviewing the iSABM-SCC decoder. Then a detailed analysis on parameter optimization of the iSABM-SCC decoder is made based on numerical simulations over an additive white Gaussian noise (AWGN) channel. The gap between the optimized iSABM-SCC decoder and the AIRs of HD-FEC is also analyzed together with the achievable transmission reach over an optical fiber channel. The results show that the optimized iSABM-SCC decoder can decrease the gap of SCCs to the AIRs of HD-FEC to $0.26$--$1.02$~dB for code rates of $0.74$--$0.87$, and increase the fiber reach by up to $22\%$ for data rates of $401$~Gbps--$468$~Gbps.

\section{System Model, Staircase Codes and the iSABM-SCC Decoder}

\subsection{System Model}

\begin{figure}[!tb]
\includegraphics[width=0.49\textwidth]{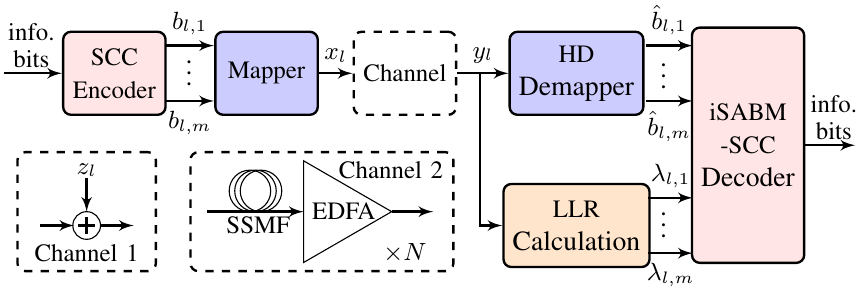}
\vspace{-2.2em}
\caption{System model considered in this paper.}
\vspace{-1.35em}
\label{fig:model}
\end{figure}

Fig.~\ref{fig:model} shows the system model. Information bits are encoded into bits $b_{l,1},\ldots,b_{l,m}$ by an SCC encoder and then are mapped to symbols $x_{l}$ taken from an equally-spaced $M$-ary pulse amplitude modulation (PAM) constellation $\mathcal{S}=\{s_{1},s_{2},\ldots,s_{M}\}$ with $M=2^m$ points, where $l$ is the discrete time index, $l=0,1,2,\ldots$. The bit-to-symbol mapping is the binary reflected Gray code. After channel transmission, an HD-demapper is used to estimate the coded bits $\hat{b}_{l,1},\ldots,\hat{b}_{l,m}$, according to the received signal $y_l$. At the same time, the receiver calculates the LLR $\lambda_{l,k}$ for each bit $\hat{b}_{l,k}$, $k=1,\ldots,m$. Both the bits $\hat{b}_{l,1},\ldots,\hat{b}_{l,m}$ and the soft information $\lambda_{l,1},\ldots,\lambda_{l,m}$ are sent to the iSABM-SCC decoder to recover the information bits. In contrast, standard SCC decoder is only fed with the bits $\hat{b}_{l,1},\ldots,\hat{b}_{l,m}$.

In this paper, the performance of iSABM-SCC is analyzed for two channels. The first one is an AWGN channel, i.e., ${y_{l}}={x_{l}}+{z_{l}}$, where ${z_{l}} \sim \mathcal{CN}(0,N_0/2)$. The signal power is normalized, i.e., $E_s\triangleq\mathbb{E}[x^2_l]=(1/M)\sum^M_is^2_i=1$, where $\mathbb{E}[\cdot]$ denotes expectation. The  signal-to-noise ratio (SNR) is defined as $\text{SNR}\triangleq E_s/N_0=1/N_0$. The LLR value $\lambda_{l,k}$ is defined as 
\begin{equation}\label{LLR}
   \lambda_{l,k}=\sum_{b \in \{0,1\}} (-1)^{\bar{b}} \log\sum_{i \in \mathcal{I}_{k,b}} \textrm{exp}\left(-\frac{(y_{l}-s_{i})^{2}}{N_0}\right),
\end{equation}
where $\bar{b}$ denotes bit negation and the set $\mathcal{I}_{k,b}$ enumerates all the constellation points in $\mathcal{S}$ whose $k$th bit is $b$.

The second channel we considered is an $N$-span optical fiber link. Each span is composed of $80$ km standard single-mode-fiber (SSMF) and an erbium-doped fiber amplifier (EDFA) with noise figure of $5$~dB. More details about the optical fiber channel will be given in Sec. III-B.

\vspace{-0.5em}
\subsection{Staircase Codes}

SCCs are built on simple algebraic component codes, e.g., Bose-Chaudhuri-Hocquenghem (BCH) codes. Let $\mathcal{C}$ be a BCH code with parameters of $(n_{c}, k_{c}, t)$, where $n_{c}$ is the codeword length, $k_{c}$ is the information length and $t$ is the error-correcting capability. An SCC based on $\mathcal{C}$ is defined as a sequence of staircase-like concatenated binary blocks $\boldsymbol{B}_{i}\in\{0,1\}^{w \times w}$, $i=0, 1, 2,\ldots$, with block size of $w=n_c/2$ and code rate of $R=2k_c/n_c-1$, where $\boldsymbol{B}_0$ is initialized to all zeros. The visualized structure of SCC is shown in~\cite[Fig.1(b)]{YiTCOM2019}. For $\forall i>1$, each row of the matrix $[\boldsymbol{B}^{T}_{i-1} \boldsymbol{B}_{i}]$ is a codeword in $\mathcal{C}$.

The decoding of SCCs is performed using a sliding window covering $L$ received blocks $\{\boldsymbol{Y}_{i}, \boldsymbol{Y}_{i+1},\ldots, \boldsymbol{Y}_{i+L-1}\}$, where $\boldsymbol{Y}_{i}$ corresponds to the transmitted block $\boldsymbol{B}_{i}$. Within the window, BDD is used to decode the rows in $[\boldsymbol{Y}^{T}_{i+L-2} \boldsymbol{Y}_{i+L-1}]$, $[\boldsymbol{Y}^{T}_{i+L-3} \boldsymbol{Y}_{i+L-2}]$,$\ldots$, $[\boldsymbol{Y}^{T}_{i} \boldsymbol{Y}_{i+1}]$ in turn, until the maximum number of iterations is reached. Then the decoder outputs $\boldsymbol{Y}_{i}$ and the window shifts to $\{\boldsymbol{Y}_{i+1}, \boldsymbol{Y}_{i+2},\ldots, \boldsymbol{Y}_{i+L}\}$ and repeats the decoding process in the last window.
To speed the decoding, $w$ BDDs can be used in parallel to decode the $w$ rows $r_1,r_2 \ldots, r_w$ of two neighbor SCC blocks. We thus treat the $w$ component decoders as a group and state that standard SCC performs $L-1$ groups of BDDs at each iteration.

Let $c_j$ be the transmitted component codeword that corresponds to the received sequence $r_j$, $j=1,\ldots,w$. The BDD output $\hat{c}_j$ is given by
\begin{subnumcases}{\hat{c}_j=}
     c_j,   & $\textrm{if~} d_{\textrm{H}}(r_j,c_j) \leq t $ \\
    \tilde{c}_j,  & $\textrm{if~} d_{\textrm{H}}(r_j,c_j) > t \textrm{~and~} d_{\textrm{H}}(r_j,\tilde{c}_j) \leq t$ \\
     r_j,  & \textrm {otherwise}
\end{subnumcases}
where $d_{\textrm{H}}(\cdot,\cdot)$ is the Hamming distance and $\tilde{c}_j$ is another codeword in $\mathcal{C}$. As (2a) shows, only when there are $t$ or less than $t$ errors, BDD can decode $r_j$ to $c_j$. Otherwise, either a miscorrection, i.e., (2b), or a failure, i.e., (2c), happens. BDD failure means that the decoder could not find a codeword in $\mathcal{C}$ that is within the Hamming distance of $t$ to $r_j$, and has to return the input $r_j$. Miscorrection is technically successful, but the output is $\tilde{c}_j$ rather than the transmitted $c_j$. Both failures and miscorrections limit the performance of BDD and thus the performance of SCCs.

\subsection{The iSABM-SCC Decoder}

To improve the performance of SCCs, the iSABM-SCC decoder was proposed in~\cite{Yi_iSABM2021}. iSABM-SCC marks bits with the help of channel LLRs to prevent miscorrections and to decode the BDD failures and miscorrections via BF. Furthermore, bit marking (BM) in iSABM-SCC uses two thresholds. Let $\delta_1$ and $\delta_2$ be the HRB threshold and HUB threshold ($\delta_1\geq\delta_2\geq0$), resp. According to the absolute value of $\lambda_{l,k}$, i.e., $|\lambda_{l,k}|$, the marking result for a bit $\hat{b}_{l,k}$ is given by
\setcounter{equation}{2}
\begin{equation}\label{2-bitMarking}
     \begin{aligned}
     \left\{
     \begin{array}{lcl}
     \text{HRB},  &      & \text{if~} |\lambda_{l,k}| \geq \delta_1 \\
     \text{UB}, &      & \text{if~} \delta_2 \leq |\lambda_{l,k}| < \delta_1 \\
     \text{HUB},  &      & \text{if~} |\lambda_{l,k}| < \delta_2\\
     \end{array}
     \right.
     \end{aligned}.
\end{equation}
A pair $(\delta_1,\delta_2)$ will be used to represent the thresholds in (\ref{2-bitMarking}).

\begin{figure}[!tb]
\centering
\includegraphics[width=0.5\textwidth]{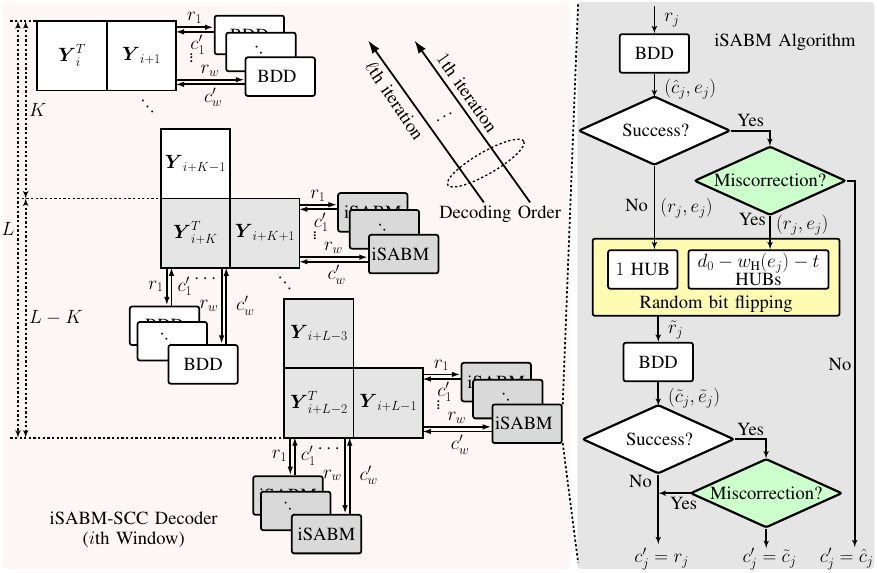}
\vspace{-2em}
\caption{Flow chart (left) of the iSABM-SCC decoder in the $i$th window. The right figure shows the workflow of iSABM to decode a component word $r_j$, $j=1,\ldots,w$. $c'_j$ is the output of the iSABM component decoder.}
\vspace{-1.6em}
\label{fig:iSABM}
\end{figure}

With the marked information in the last $L-K$ blocks of a window, $K=0,\ldots,L-2$, iSABM-SCC performs $K$ groups of BDDs and $L-K-1$ groups of iSABMs at each iteration. This is shown in the left side of Fig.~\ref{fig:iSABM}.

The right side of Fig.~\ref{fig:iSABM} shows the workflow of iSABM to decode a received sequence $r_j$. Similar to standard SCC decoding, BDD is performed first. In the event of success, iSABM will check whether the BDD output $\hat{c}_j$ is a miscorrection or not. The criteria are that the detected errors should not be in conflict with (i) HRBs or (ii) any bit in a correctly decoded component codeword. Only when the two criteria are satisfied, the output of BDD will be accepted, i.e., $c'_j=\hat{c}_j$. Otherwise, $\hat{c}_j$ will be regarded as a miscorrection and be rejected.

For the miscorrections, the iSABM algorithm \emph{randomly} selects $d_{0}-w_{\text{H}}(e_j)-t$ HUBs in $r_j$ for flipping, where $d_0$ is the minimum Hamming distance between the codewords in $\mathcal{C}$ and $w_{\text{H}}(e_j)$ is the Hamming weight of the error pattern $e_j$ detected by BDD. In the case of BDD failure, the iSABM algorithm \emph{randomly} selects $1$ HUB in $r_j$ for flipping. The intuition here is that in some cases, BF will make the resulted sequence $\tilde{r}_j$ close enough to the transmitted codeword $c_j$, i.e., $d_\text{H}(\tilde{r}_j,c_j)=t$. Thus when BDD is excuted again, the residual errors in $\tilde{r}_j$ can be corrected. Similarly, successful BDD case follows a step of miscorrection detection in case of miscorrections caused by BF. In addition, in some cases the number of HUBs is less than the required number of bit flips ($1$ for failure recovery and $d_{0}-w_{\text{H}}(e_j)-t$ for miscorrection recovery). If
that happens, even if all the HUBs are flipped, BF can not bring $\tilde{r}_j$ close enough to $c_j$. Hence, the iSABM decoding will stop and output $r_j$, and then proceed to decode the next component word.

In the iSABM-SCC decoder, there is no need to store LLRs. iSABM-SCC only requires $2$ bits to represent the three groups: HRB, HUB and UB. The marked information is not updated either. To reduce the memory requirement, $1$-bit marking is analysed as well. Let $\delta_3$ be the $1$-bit marking threshold, the marking result for a bit $\hat{b}_{l,k}$ is given by
\begin{equation}\label{1-bitMarking}
     \begin{aligned}
     \left\{
     \begin{array}{lcl}
     \text{HRB},  &      & \text{if~} |\lambda_{l,k}| \geq \delta_3 \\
     \text{HUB},  &      & \text{if~} |\lambda_{l,k}| < \delta_3\\
    \end{array}
     \right.
     \end{aligned}.
\end{equation}

\section{Simulation Results}

\begin{figure}[!tb]
\centering
\includegraphics[width=0.45\textwidth]{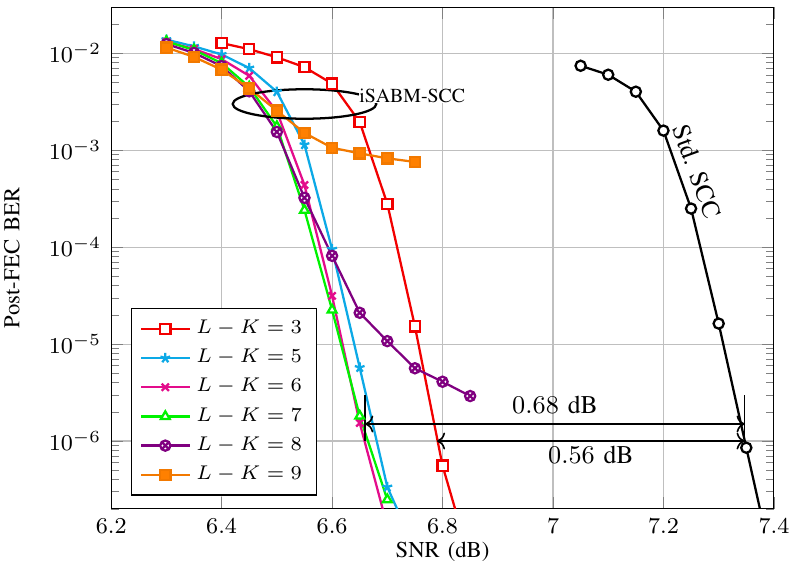}
\vspace{-1em}
\caption{BER performance of iSABM-SCCs based on $\mathcal{C}_1$ under different $L-K$ values for 2-PAM.}
\label{fig:L-K_Optimization}
\vspace{-0.5em}
\end{figure}

Table~\ref{tab:SCCs} shows the SCC parameters we used. Let $\nu$ be an integer, the codeword length and information length of $\mathcal{C}$ are given by $n_c=2^{\nu}-1+u$ and $k_c=n-\nu t-u$, resp., where $u$ is the number of extended parity bits. The number of iterations is $\ell=7$, while the decoding window size is $L=9$.
\vspace{-1em}
\begin{table}[t]
\caption{SCC parameters we used in this paper}
\centering
{
\begin{tabular}{cccccccccc}
\hline

\hline
& $R$ & $\nu$ & $u$ & $t$ & $n_c$ & $k_c$ & $w$\\
\hline

\hline
$\mathcal{C}_1$ & 0.87 & 8 & 1 & 2 & 256 & 239 & 128\\
$\mathcal{C}_2$ & 0.80 & 8 & 1 & 3 & 256 & 231 & 128\\
$\mathcal{C}_3$ & 0.74 & 8 & 1 & 4 & 256 & 223 & 128\\
\hline

\hline
\end{tabular}
}
\label{tab:SCCs}
\vspace{-1em}
\end{table}

\vspace{0.5em}
\subsection{Parameter Optimization}

To maximize the performance of iSABM-SCC, it is key to carefully select the values of the parameters $L-K$ as well as $(\delta_1,\delta_2)$ for (3) or $\delta_3$ for (4). In what follows, we will discuss the optimization of the parameters $L-K$, $(\delta_1,\delta_2)$ and $\delta_3$ using numerical simulations over an AWGN channel.

Fig.~\ref{fig:L-K_Optimization} shows the performance of iSABM-SCCs based on $\mathcal{C}_1$ and 2-PAM for different $L-K$ values. BM uses (\ref{2-bitMarking}) with thresholds of $(10,2.5)$. Fig.~\ref{fig:L-K_Optimization} shows that for $L-K \leq 7$, the gain of iSABM-SCC with respect to standard SCC increases along with $L-K$, and achieves a maximum value of $0.68$~dB at a post-FEC bit-error ratio (BER) of $10^{-6}$ for $L-K=6, 7$.
Once $L-K>7$, an error floor appears. The reason for this is as follows. For the sliding window fashion, the blocks that are closer to the top left of the window are decoded more times. Hence, there is little probability of miscorrection for the BDDs related to the first two blocks at the top left of the window (due to the few errors). However, due to the inaccurate marked HRBs, not all miscorrections are identified. The mis-regarded correctly decoded component codewords and the inaccurate HRBs together tend to lead to the rejection of the correct BDDs related to the first two blocks of the window, and thus cause serious performance loss.

\begin{figure*}[!tb]
\centering
\includegraphics[width=0.945\textwidth]{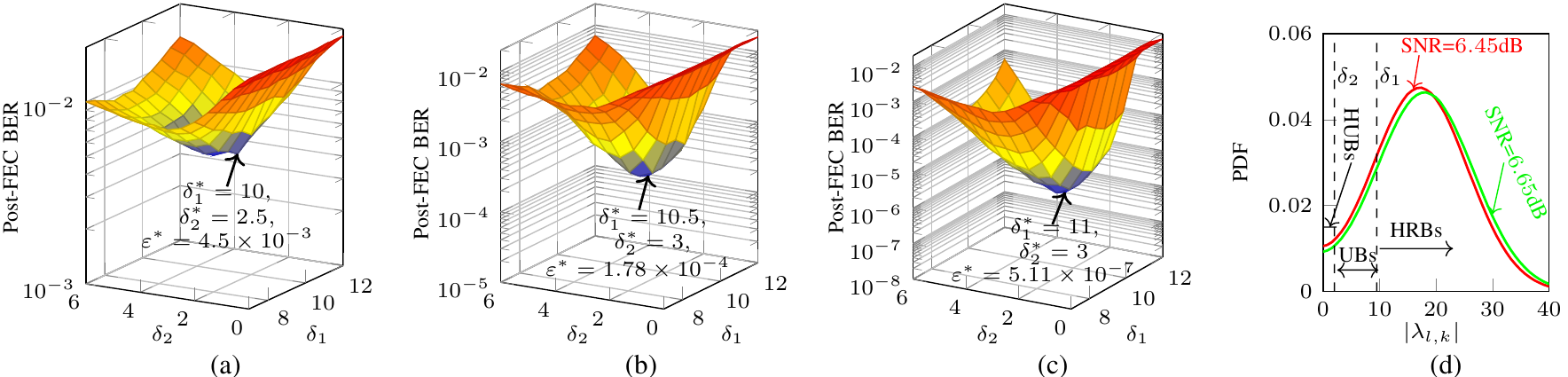}
\vspace{-0.9em}
\caption{Optimization of $(\delta_1,\delta_2)$ for iSABM-SCC based on $\mathcal{C}_1$ for 2-PAM and SNRs of (a) $6.45$~dB, (b) $6.55$~dB and (a) $6.65$~dB. The rightmost figure (d) is the PDFs of $|\lambda_{l,k}|$ for SNRs of $6.45$~dB and $6.65$~dB. The optimal post-FEC BER is shown with $\varepsilon^*$. }
\vspace{-1.2em}
\label{fig:delta12Optimization}
\end{figure*}

Fig.~\ref{fig:delta12Optimization}(a)--(c) shows the optimization of $(\delta_1,\delta_2)$ for iSABM-SCC based on $\mathcal{C}_1$ with $L-K=7$ and (\ref{2-bitMarking}) for BM. They are tested for SNRs of $6.45$~dB, $6.55$~dB and $6.65$~dB, which are picked from the waterfall region shown in~Fig.~\ref{fig:L-K_Optimization}. In general, the cone-like shape indicates that whenever the values of $\delta_1$ and $\delta_2$ are too large or too small, the decoding performance of iSABM-SCC will degrade. If $\delta_1$ is too large, the HRB class may miss some highly reliable bits that weakens the ability of the decoder to prevent miscorrections. On the contrary, a small $\delta_1$ tends to mistakenly mark some errors as HRBs. As a result, some correct BDD outputs might be mistakenly interpreted as miscorrections and be rejected by the decoder. In the case of iSABM-SCC with a small $\delta_2$, the number of HUBs reduces, which increases the probability that there is no enough HUBs for flipping. On the contrary, a large $\delta_2$
may cause that some correct bits are mistakenly marked as HUBs. This increases the probability that the flipped bits are not errors and thus causes failures or miscorrections in the second BDD trial. All the mentioned above potentially give a performance deterioration.

Fig.~\ref{fig:delta12Optimization}(a)--(c) also shows that the optimal value of $(\delta_1,\delta_2)$, denoted by $(\delta^*_1,\delta^*_2)$, are $(10,2.5)$, $(10.5,3)$ and $(11,3)$ for SNRs of $6.45$~dB, $6.55$~dB and $6.65$~dB, resp. The corresponding optimal post-FEC BERs (denoted by $\varepsilon^*$) are $4.5\times10^{-3}$, $1.78\times10^{-4}$ and $5.11\times10^{-7}$. It is observed that the value of $(\delta^*_1,\delta^*_2)$ increases as SNR increases. Intuitively, this is due to that the decoder wants to resist the change of the proportion of HRBs, UBs and HUBs in the total bits, as the distribution of $|\lambda_{l,k}|$ shifts.

Fig.~\ref{fig:delta12Optimization}(d) shows how the probability density function (PDF) curve of $|\lambda_{l,k}|$ shifts to the right when SNR increases from $6.45$~dB to $6.65$~dB. The integrals between the PDF curve and the $x$-axis over the intervals $[0,\delta_2)$, $[\delta_2,\delta_1)$ and $[\delta_1,\infty)$ are the proportion of HUBs, UBs and HRBs in the bits, resp. Due to the right shift of the PDF curve, the proportion of HUBs and UBs decreases, while that of HRBs increases, if the value of $(\delta_1,\delta_2)$ is fixed, e.g., $(10,2.5)$. Increasing the thresholds to the optimal one, i.e., $(11,3)$ for SNR of $6.65$~dB, will increase the proportion of HUBs and UBs and decrease the proportion of HRBs. This shift of the optimal thresholds seems to want to keep the proportion of HRBs, UBs, and HUBs constant. To ensure the best performance for all SNRs, this may need to be considered in future work.

\begin{figure}[!tb]
\centering
\includegraphics[width=0.44\textwidth]{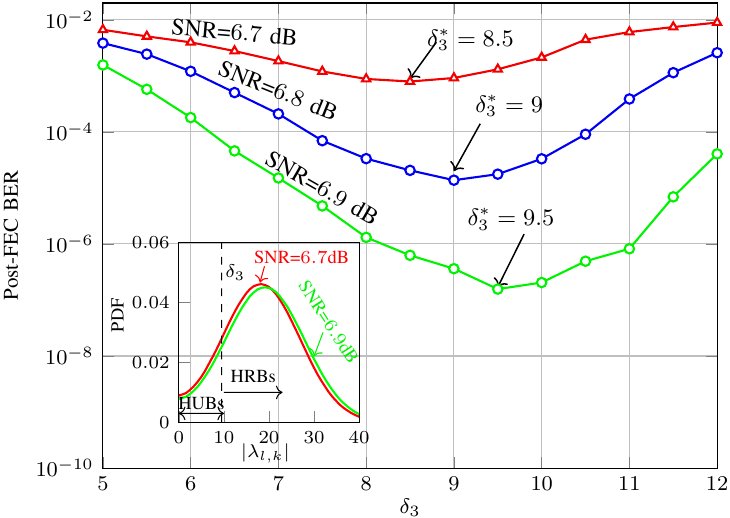}
\vspace{-1em}
\caption{Optimization of $\delta_3$ for iSABM-SCC based on $\mathcal{C}_1$ for 2-PAM. The inset figure is the PDFs of $|\lambda_{l,k}|$ for SNRs of $6.7$~dB and $6.9$~dB.}
\vspace{-1.5em}
\label{fig:delta3Optimization}
\end{figure}

Fig.~\ref{fig:delta3Optimization} shows the optimization of $\delta_3$ for iSABM-SCC based on $\mathcal{C}_1$ using (\ref{1-bitMarking}) for BM and $L-K=7$. The three curves were studied for SNRs of $6.7$~dB, $6.8$~dB and $6.9$~dB, while the inset is the PDFs of $|\lambda_{l,k}|$ at SNRs of $6.7$~dB and $6.9$~dB. The corresponding optimal thresholds $\delta^*_3$ are found to be $8.5$, $9$ and $9.5$. On the one hand, the U-type curve indicates that the value of $\delta_3$ should not be too large or too small, either. On the other hand, a similar trend as that shown in Fig.~\ref{fig:delta12Optimization} is observed, i.e., the optimal marking threshold increases as SNR increases.

\vspace{-0.35em}
\subsection{Performance Analysis}
\vspace{-0.32em}

\begin{figure}[!tb]
\centering
\includegraphics[width=0.455\textwidth]{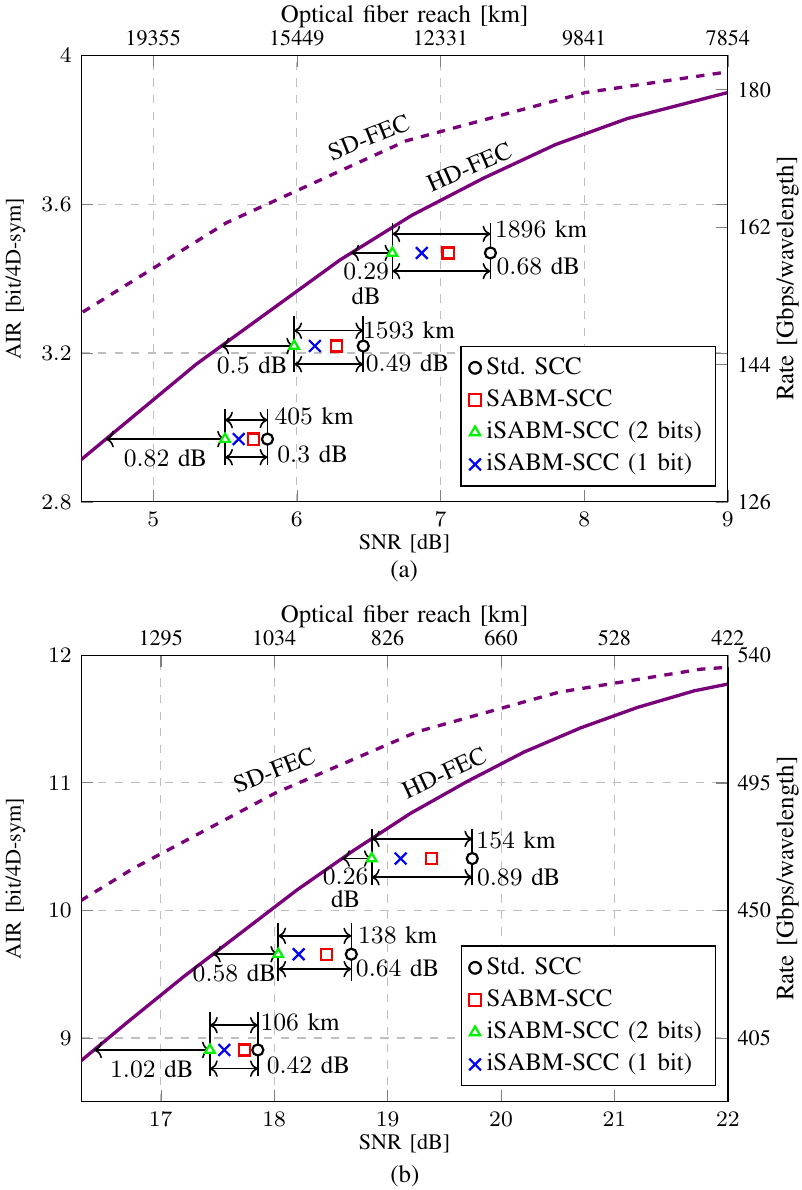}
\vspace{-1em}
\caption{AIRs for HD-FEC and SD-FEC with dual-polarization I/Q-modulated (a) 2-PAM ($m=1$) and (b) 8-PAM ($m=3$). Markers show the results for iSABM-SCC, SABM-SCC and standard SCC.}
\vspace{-1.6em}
\label{fig:AIR}
\end{figure}

Fig.~\ref{fig:AIR} shows the required SNRs for iSABM-SCC (with $L-K=7$), SABM-SCC and standard SCC to achieve a post-FEC BER of $10^{-6}$  ($x$-axis at the bottom).  The rates are calculated as $I_\text{SCC}=4mR$ with $R\in\{0.74,0.80,0.87\}$, where the factor of $4$ represents four dimensions (4D) of dual-polarization and in-phase-quadrature (I/Q) of the optical signal. For iSABM-SCC using (\ref{2-bitMarking}), the marking thresholds we used are $(\delta_1,\delta_2)=(10,2.5)$, which are optimized for iSABM-SCC based on $\mathcal{C}_1$ at SNR of $6.45$~dB shown in Fig.~\ref{fig:delta12Optimization}(a). For iSABM-SCC using (\ref{1-bitMarking}), we simply set $\delta_3=\delta_1$, i.e., $\delta_3=10$, which gives a very close performance to iSABM-SCC with the optimal threshold at each tested SNR shown in Fig.~\ref{fig:delta3Optimization}. The thresholds we used in Fig.~\ref{fig:AIR} are fixed for all SNRs and code rates (this is also how it was done in~\cite{Yi_iSABM2021}). Fig.~\ref{fig:AIR} shows that iSABM-SCC can achieve $0.3$--$0.68$~dB ($0.42$--$0.89$~dB) extra gain for 2-PAM (8-PAM), when compared to standard SCCs. The performance of iSABM-SCC is significantly better than that of SABM-SCC, even with 1-bit marking.

The solid and dashed curves in Fig.~\ref{fig:AIR} are the maximum AIRs of HD-FEC and SD-FEC, resp. The AIR of SD-FEC is the generalized mutual information~\cite[eq.(14)]{AIR_AlexJLT2017}, while that of HD-FEC is given by $I_\text{HD}  = 4m (1+p\log_2p+(1-p)\log_2(1-p))$, where $p$ is the average pre-FEC BER across the $m$ bits mapped into the selected single-dimensional modulation format. Fig.~\ref{fig:AIR} shows that due to the finite block length and limited complexity, there is a gap between the practical SNRs required for SCCs and the predicted SNRs via AIR. By replacing standard decoding with iSABM-SCC decoding, the gap of SCCs to the AIR preditions can be reduced to $0.29$--$0.82$~dB ($0.26$--$1.02$~dB) for 2-PAM (8-PAM). However, note that due to the use of soft information, the AIR of HD-FEC is not a true upperbound for the SA-HD decoder. An AIR upperbound for the SA-HD decoders is expected to be located between the solid and dashed curves. Accurately estimating such an upperbound is still an open problem.

Fig.~\ref{fig:AIR} also shows the achievable rate and optical transmission reach for the SCC decoders ($x$-axis at the top). A dual-polarization wavelength-division-multiplexing coherent communication system is considered with $45$ GBaud symbol rate, $11$~channels, $50$~GHz channel spacing and $1550$~nm center wavelength. The optical fiber is with $0.2$ dB/km loss and $17$~ps/nm/km dispersion. The nonlinear coefficient of the fiber is $1.2$ (W$\cdot$km)$^{-1}$. The reach increase is evaluated based on the Gaussian noise model proposed in~\cite{GNmodel2014}.
Fig.~\ref{fig:AIR} shows that the achieved extra gains of iSABM-SCCs (with respect to standard SCCs) yields reach increases of up to $1896$~km ($17\%$) for data rates between $134$~Gb/s and $156$~Gb/s, and up to $154$~km ($22\%$) for data rates between $401$~Gb/s and $468$~Gb/s per wavelength.

\section{Conclusions}
In this paper, two aspects of the improved soft-aided bit-marking staircase decoder (iSABM-SCC) were analyzed. The first one is a parameter optimization for iSABM-SCC, while the second one is an analysis of the gap of iSABM-SCC to the AIR of HD-FEC in an AWGN channel and the reach enhancement in the optical fiber channel. It was shown that carefully optimizing the marking thresholds and the number of modified component decodings is key for iSABM-SCC to provide the best performance.

\bibliographystyle{IEEEtran}

\bibliography{refs}

\end{document}